\documentclass[12pt,preprint]{aastex}

\slugcomment{Accepted by the Astrophysical Journal}
\shorttitle{X-ray Desorption of Molecules from Grains}
\shortauthors{Najita, Bergin, \& Ullom}

\newcommand{\nH}{\rm n_H}
\newcommand{\Msun}{M_\odot}
\newcommand{\ergpers}{\rm \,erg\,s^{-1}}
\newcommand{\gpersqc}{\rm \,g\,cm^{-2}}
\newcommand{\AU}{\rm \,AU}
\newcommand{\degK}{\rm \,K}
\newcommand{\eV}{\rm \,eV}

\begin{document}

\title{X-ray Desorption of Molecules from Grains in Protoplanetary Disks}
\author{Joan Najita}
\affil{NOAO, 950 N. Cherry Ave, Tucson, AZ 85719}
\email{najita@noao.edu} 
\author{Edwin A. Bergin} 
\affil{Harvard-Smithsonian Center for Astrophysics, MS 66, 60 Garden St., 
Cambridge MA, 02138} 
\email{ebergin@cfa.harvard.edu} 
\author{J.N. Ullom}
\affil{Lawrence Livermore National Laboratory, 
7000 East Ave., MS L418, Livermore CA, 94550}
\email{ullom1@llnl.gov}

\begin{abstract} 
We examine the ability of stellar X-rays to desorb molecules from
grains in outer protoplanetary disks. In particular, we consider the
possibility of spot heating by X-rays and examine its effectiveness,
compared to whole grain heating by X-rays, in sustaining a gas phase
abundance of CO.  As found in previous studies, whole grain heating is
effective only for small grains ($< 500$\AA).  As a result, large
grains are a permanent sink of CO and whole grain
heating cannot sustain an equilibrium gas phase abundance of CO.  Spot
heating, in which the incident X-ray deposits energy in only a
restricted region of the grain volume and which may occur as a
consequence of the aggregate nature of grains, proves to be more
promising.  Assuming that grains are comprised of many thermal subunits
that are poorly connected thermally, we find that spot heating is
efficient at all grain sizes if large grains are effectively ``coated''
with small grains $\lesssim 100$\AA\ in size.  We discuss the implications
of X-ray spot heating for the interpretation of millimeter emission
line studies of outer protoplanetary disks.

\end{abstract}


\section{Introduction}

An understanding of the gas phase abundances of outer protoplanetary disks 
($r>100$ AU) is important for our understanding of nebular chemistry 
and for the interpretation of millimeter and sub-millimeter emission line 
studies of disks.
Among the important processes at these radii 
is the desorption of molecules from grains. 
At the the high density ($\nH > 10^6$ cm$^{-3}$), 
low temperature conditions characteristic of these regions, 
molecules are expected to adsorb readily onto grains, 
making it difficult for molecules to remain in the gas phase   
in the absence of efficient desorption mechanisms. 
Millimeter wave measurements of disks strongly hint at the efficacy of 
molecular depletion. 
For example, CO is one of the best diagnostics of outer disks due to 
its low condensation temperature ($\sim 20$ K), chemical robustness, and 
consequent high expected abundance in the absence of depletion.
Studies at millimeter wavelengths indeed find that CO emission from 
outer disks is much weaker than would be predicted 
from disk continuum emission strengths in the absence of depletion.  
These results imply that a large fraction of CO (95$-$99\%) 
is depleted onto grains (e.g., Dutrey et al. 1996). 
Currently, molecular emission is detected from only a few systems, with the 
emission typically extending over $\sim$500 AU.

It is, perhaps, surprising that there is a detectable gas phase 
in these systems. 
At disk radii $> 50$ AU from the typical T Tauri star,  
grain temperatures from the disk photosphere to the disk midplane are 
below $\sim 20$K (Beckwith et al.\ 1990; 
D'Alessio et al. 1998).  As a result, grains at these radii 
(except those in the upper disk photosphere) 
are too cool to contribute significantly to the thermal desorption of 
even molecules such as CO.  
The expected rapid depletion of CO from the gas phase at these radii 
is borne out in detailed disk chemistry calculations.  For example, 
Aikawa et al.~(1996) explored the depletion of CO in protoplanetary 
disks at large radii ($> 100$ AU) assuming low grain temperatures and 
no non-thermal desorption mechanisms.  They found that at a disk 
radius of 200 AU, CO is depleted by a few orders of magnitude 
below values typical of the ISM on a time scale of $10^5-10^6$ yr.  
Since much stronger CO emission lines are observed than are predicted 
by these models, 
Aikawa \& Herbst (2001) have more recently argued for the role of 
some kind of non-thermal desorption mechanism in order to explain the 
line strengths observed in systems such as DM Tau.

A possibly significant source of desorption is 
stellar irradiation at optical and X-ray wavelengths. 
The effect of stellar irradiation at optical wavelengths 
on the vertical temperature 
structure of disks has been considered previously 
(e.g., D'Alessio et al.\ 1998; Calvet et al.\ 1991; 
Chiang \& Goldreich 1997).  These authors demonstrate that even at 
100 AU distances, grains at irradiated disk surfaces will be heated 
to temperatures in excess of the disk effective temperature due, 
in part, to the mismatch in the optical absorption and far-infrared 
emission efficiencies of grains.  This may produce thermal 
desorption at the disk surface. 

In the present study, we consider the possible role of stellar 
X-rays as an additional, non-thermal desorption mechanism.  
As is well-known, pre-main-sequence stars are strong emitters of 
stellar X-rays, with X-ray emission at 10 to $>10^4$ times 
typical main-sequence levels (Feigelson \& Montmerle 1999).
X-ray absorption may be a useful desorption mechanism since 
the penetration depth of X-rays is larger than that of 
optical and UV photons.  At the standard gas-to-dust ratio, the 
absorption optical depth for 1.5 keV photons 
(e.g., Morrison \& McCammon 1983) is comparable to the 
optical depth for dust absorption at $\sim 2\mu$m.  
Whole grain heating by X-rays has been considered previously 
as a desorption mechanism in the 
context of molecular clouds (e.g., Leger et al.\ 1985). 
Desorption due to localized, transient heating events 
(``spot heating'') has been explored in the context of 
grain heating by cosmic rays (Leger et al.\ 1985) and by 
photons at ultraviolet through infrared wavelengths (Duley et al.\ 1989). 
Here, we extend the Leger et al.\ study of X-ray desorption using the 
grain heating calculation of Dwek \& Smith (1996) and consider 
in greater detail the possibility of spot heating by X-rays.

\section{Molecular Desorption by Stellar X-rays}
The molecular desorption rate is governed by the rate of X-ray energy 
deposition into the grain.  For a grain of radius $a$, the photon 
absorption rate ($A_X$), energy deposition rate ($H_X$),
and molecular desorption rate ($D_X$) are  
\begin{equation}
A_X (a) = \int_{E_{\rm min}}^{E_{\rm max}} F_X(E)\ \sigma_X(a,E)\ dE,  
\end{equation}
\begin{equation}
H_X (a) = \int_{E_{\rm min}}^{E_{\rm max}} F_X(E)\ \sigma_X(a,E)\ E_d(a,E)\ dE
\end{equation}
\begin{equation}
D_X (a) = \int_{E_{\rm min}}^{E_{\rm max}} F_X(E)\ \sigma_X(a,E)\ N_{\rm des} (a,E)\ dE
\end{equation}
where $F_X$ is the specific photon (number) flux of X-rays of energy $E$ 
that is
incident on the grain, $\sigma_X$ is the effective X-ray absorption
cross-section for the grain, $E_d$ is the energy deposited
in the grain as heat ($E_d \le E$), and 
$N_{\rm des}$ is the number of molecules desorbed by an incident photon 
of energy $E$.  The effective cross-section, deposited energy, and 
desorption rates are
discussed in more detail below. 

\subsection{Stellar X-ray Flux}
The stellar X-ray emission is assumed to be thermal bremsstrahlung with 
a spectrum of the form
\begin{equation}
L_{\nu\ast}= L_{\nu 0} \exp(-E/T_X) 
\end{equation}
where $T_X$ is the characteristic X-ray plasma temperature and 
$L_{\nu 0}$ is a normalization constant. 
{\it ASCA\/} and {\it ROSAT\/} studies of young stars generally 
find that $T_X \simeq 1-2$~keV and that the stellar X-ray flux 
is a significant fraction of the stellar bolometric luminosity. 
A fiducial value is $L_X/L_\ast \simeq 10^{-4}$, although more 
recent results from {\it Chandra} on the Orion Nebula Cluster 
find that young $\approx 1 \Msun$ stars have integrated X-ray 
luminosities of $L_X \sim 2\times 10^{30}\ergpers$ between 
0.2 keV and 2 keV (Garmire et al.\ 2000). 
In the present study, we set the normalization constant $L_{\nu 0}$ 
using the mean X-ray properties of Garmire et al.\ (2000). 

Grains suspended in the disk atmosphere at large distances from 
the stellar X-ray source will see an X-ray photon flux of 
$F_X = L_{\nu\ast}e^{-\tau_\nu}/4\pi d^2 E$, 
where the specific optical depth to the X-ray source, $\tau_\nu,$ 
includes the contribution from overlying disk layers.  
In this expression, we have assumed that the observed X-ray flux 
includes only the half emitted by the hemisphere facing 
the observer.  The emission from the other hemisphere is assumed to be 
occulted by the disk but is available to heat its corresponding disk 
surface. 
As an approximation, the radiative transfer of the X-rays is treated 
as pure attenuation. 

We assume that the disk has the column density of the minimum mass 
solar nebula 
$\Sigma_D = 1500 \gpersqc (r/1\AU)^{-3/2}.$
The disk is also assumed to be vertically isothermal and in vertical 
hydrostatic equilibrium. 
Following D'Alessio et al. (1998), 
we assume a disk temperature of $T_0=17$ K at 100 AU; 
$T_D(r) = T_0 (r/100\AU)^{-1/2}$ within 100 AU; and 
$T_D = T_0$ beyond 100 AU. 
We integrate numerically over the resulting disk density distribution 
in order to determine the line-of-sight column density from the X-ray 
source to a given point in the disk.

\subsection{Heat Deposition by X-rays}
Beyond the stellar X-ray flux, we also need to determine $\sigma_X$,
the grain absorption cross-section, and $E_d$, the energy deposited per
absorbed photon, in order to determine the resulting rate of molecular
evaporation.  We used the formalism of Dwek \& Smith (1996) to calculate
the X-ray absorption cross-section $\sigma_X$ for spherical grains
averaged over photon impact parameter (see their equations 1, 2).
Here the absorption cross-section is expressed as 
$\sigma_X = \pi a^2 P_{\rm abs}$, 
the product of the geometric cross-section and an absorption probability.

The details of the absorption process are described by Dwek \& Smith (1996). 
In the absorption process, 
the absorption of the X-ray results in a hot primary and Auger 
electrons.  These interact with the solid lattice on their way out of 
the grain, losing energy which goes into grain heating.
Only a fraction of the incident photon energy is deposited in 
the grain because some of the electrons escape with 
finite energy.
Thus, as discussed previously by Leger et al.\ (1985), there is 
an optimal grain size range in which most of the photon energy 
goes into heating the grain to an appreciable temperature.  
On the one hand, very small grains have a small probability of 
absorption due to their limited thickness, and a lower probability 
for the liberated electron energy to be entirely deposited 
in the grain for the same reason.  
But when photons are absorbed, the small grain is heated to high 
temperature.  On the other hand, very large grains have 
an absorption probability of unity, but are difficult to heat 
to high temperature due to their large volume and their correspondingly 
higher heat capacity. 

To determine $E_d$,   
we used the tabulation by Dwek \& Smith (1996) of 
the product $P_{\rm abs}\ E_d$ (their Table 5). 
To use their tabulated values, we calculated $P_{\rm abs}$ by first 
fitting the cross-sections of Henke et al. (1993) for the individual 
elements that make up a silicate grain, and  
then summed them as appropriate for a silicate grain 
(i.e., (Mg Si Fe)O$_4$ as in Dwek \& Smith 1996). 
We then calculated  $P_{\rm abs}$ for the tabulated grain sizes 
and incident photon energies, and finally 
divided the tabulated quantities by the calculated values of 
$P_{\rm abs}$ to obtain the deposited energies.
The results (Figure 1) show that 0.5--5 keV X-rays 
(i.e., energies 
relevant for X-ray emission from young stars) deposit a reasonable 
amount of energy (0.3--1.6 keV) in grains 50--700\AA\ in size 
(see also figure 5 of Dwek \& Smith 1996).  
In subsequent calculations, we used linear interpolation 
of the resulting table of deposited energies 
to obtain $E_d$ for each silicate grain size.

\subsection{Whole Grain Heating}
Given the deposited energy as a function of 
grain radius $a$ and incident photon energy $E$, 
we can now calculate the number of 
molecules that are evaporated as a function of $a$ and $E$ 
assuming that the absorbed energy 
is shared throughout the entire grain.  
For the heat capacity of silicate grains,  
the final grain temperature $T_f$ can be quite high, 
especially for smaller grain sizes. 
The heat capacity of the grain per unit volume is 
$\rho C_V = 1.4\times 10^2\ T^2$ J m$^{-3}$ K$^{-1}$ at low 
temperatures, $T=10-50$ K, and 
$\rho C_V = 2.2\times 10^3\ T^{1.3}$ J m$^{-3}$ K$^{-1}$ at 
higher temperatures, $T=50-150$ K  
(e.g., Leger et al.\ 1985; see also Appendix A).  
With these values, 
grains 50\AA\ in size are heated to $>120\degK$ by the 
absorption of photons with energies $\gtrsim $300~eV, and  
300\AA\ grains are heated to $>30\degK$ by the 
absorption of photons with energies 1350$-$4500~eV. 
Grains larger than 500\AA\ cannot be heated above $20\degK$ by the 
absorption of photons of any energy. 
The heated grains will evaporate molecules down to a temperature 
of $T_r \simeq 25$~K, below which they will cool predominantly by 
radiation rather than evaporation (e.g., Leger et al.\ 1985).  
As a result, only a fraction of the deposited energy is 
available for desorption. 
The energy that is available for molecular desorption is 
approximately 
\begin{equation}
E_{\rm des} = E_d 
	  - 1.26 \eV\ 
	    a_{1000}^3
	    \left(T_r^3 - T_i^3\right)
\end{equation}
where $a_{1000}$ is the grain size in units of 1000\AA, 
and $T_i$ is the initial grain temperature in degrees K. 
Since the energy required to desorb a single CO molecule is 0.083 eV, 
the number of molecules that can be evaporated is
$N_{\rm des} = E_{\rm des}/0.083\eV$.
At a given disk radius, the desorption rate will decrease at larger 
vertical column densities due to the increased disk column density 
along the line of sight to the X-ray source.  At the same time, 
the higher density at larger vertical column densities produces an 
increased molecular adsorption rate.  These two trends conspire to 
produce a sharp transition from a surface layer where X-ray desorption 
is competitive with adsorption to lower layers where X-ray desorption 
cannot compete with adsorption. 

A comparison of the desorption rate estimated above with fiducial 
adsorption rates demonstrates the limited power of whole grain heating.
To estimate the adsorption rate for the disk parameters given 
in \S 2.1, we assume a total carbon abundance of $3.3\times 10^{-4}$ 
relative to H, that half of all carbon is in the gas phase as CO, 
and that the gas and dust are fully mixed. 
The adsorption rate of a dust grain of radius $a$ in the 
upper disk atmosphere is then 
$R_{\rm ads} = \pi a^2 n v S$
where $S=0.3$ is the sticking probability (Williams 1993), 
and $n$ and $v$ are the number density and thermal speed of CO.

Comparing the X-ray desorption and thermal adsorption rates, 
we find that X-ray desorption is competitive with adsorption of CO 
for small grains over modest disk column densities.  For example, 
for $300$\AA\ grains, X-ray desorption is competitive with adsorption 
over vertical column densities of 
approximately $0.01 \gpersqc$ at disk radii of 100--300AU, 
where equal contributions are assumed from the top and bottom 
illuminated surfaces of the disk.  
However, a CO gas phase cannot be retained indefinitely over the 
same column density since CO cannot be desorbed from grains 
$\gtrsim 500$ \AA\ if whole grain heating is the only desorption mechanism. 
In fact, since larger grains are a permanent sink of gas phase CO, with 
only whole grain heating, the gas phase abundance of CO would disappear 
at all scale heights on short timescales.  
If the temperature is $< 20$ K, in
regions of the disk atmosphere at densities $\sim 10^7$ cm$^{-3}$,  
gas-grain chemical models with an MRN distribution of grain sizes 
show that the e-folding depletion timescale is $600$ yr.  
In the mid-plane at 100 AU, the depletion timescales are much shorter, 
$< 1$ yr!
Longer timescales would result if the grain size distribution 
is much steeper than MRN.  This situation might characterize the 
upper disk atmosphere as a result of the grain sedimentation process.  
Even under these circumstances, however, the effect is relatively 
small.  For example, for models with a grain size distribution 
$n(a) \sim a^{-4.5}$, 
the e-folding depletion time scale is still short, $\sim 1200$ yr.

\section{Spot Heating}

X-ray desorption would be more effective at maintaining a gas phase if 
it could also desorb molecules from large grains.  This would be 
possible, for example, if X-rays heat only small portions of large grains.  
The deposition of heat 
in a small volume could heat that portion of the grain to high 
temperature, resulting in significant desorption, as long as  
the desorption occurs rapidly compared to the time for heat to 
diffuse out of the small volume (``spot heating'').  

A natural way in which this might occur is as a consequence of 
grain growth through coagulation, a process that is expected to 
produce porous, fluffy aggregates.  
The concept and effects of a fluffy grain structure on the basic grain
properties, gas-grain dynamics, and dust extinction have received much
attention in the literature (e.g. Draine 1985; Meakin \& Donn 1988;
Mathis \& Whiffen 1989; Ossenkopf 1993; Ossenkopf \& Henning 1994;
Dominik \& Tielens 1997; Witt 2000).  In the denser regions of the ISM, 
the lower extinction per hydrogen nucleus 
(Cardelli, Clayton, \& Mathis 1989) and the implied 
larger size of scattering grains (Tielens 1989), suggests a larger 
average grain size in these regions, which probably results from   
the coagulation of smaller particles.  In protoplanetary disks, 
the sedimentation of larger grains to the disk midplane allows for the
origin of macroscopic solids through the coagulation of numerous smaller
bodies which are themselves composite structures (e.g. Weidenschilling
\& Cuzzi 1993).  However, except in the context of CO desorption from
hydrogenated amorphous carbon (Duley \& Williams 1988), the effects
of grain sub-structure have not been examined in the context of the
gas-grain adsorption and desorption.

In our model of a silicate aggregate we draw upon the literature
describing the physics of the coagulation process (Chokshi et al. 1993;
Ossenkopf 1993; Dominik \& Tielens 1997).  From the point of view of heat
transfer through a microscopic aggregate composed of conjoined small
particles, the most important aspect is the size of contact area which
allows heat to flow from one portion of the aggregate to the next.
As described by Chokshi et al. (1993), when 
two particles of size $r_s$ conjoin, they meet over a small but finite 
contact area of radius $r_c$.  In their study of 1000\AA\ grains, 
$r_c/r_s = 0.1$ for ``sticky'' 
materials such as water ice, and $r_c/r_s=0.02$ for less ``sticky'' 
materials such as quartz.  We adopt these ratios in our study. 
The limited contact area, and the imperfect conductivity across the 
contact area, may produce small grain volumes that are relatively 
thermally isolated, and therefore prone to spot heating.

In order to estimate the efficiency of spot heating that might 
result in this situation, we 
modeled grains as comprised of many thermal subunits that are poorly 
connected thermally and explored the effectiveness of X-ray spot heating 
when cooling by molecular evaporation competes with radiative cooling 
and thermal diffusion in the grain.
We approximated the thermal structure of a large grain 
of radius $a$ as a linear chain of many, $N_{s}$, smaller 
approximately spherical thermal subunits of radius $r_{s}$. 
When one of the thermal subunits, e.g., on the surface of 
the grain, is hit by an X-ray, a fraction of the incident photon 
energy will be deposited in the subunit.  For example, if 
$r_s=50$ \AA, then $\gtrsim 300$ eV 
will be deposited in the subunit for incident photons in the 
energy range $\sim 800-3500$ eV (\S 2.2). 

If we assume that the deposited energy is immediately thermalized 
in the thermal subunit, then the equation that governs how the 
thermalized energy diffuses through the rest of the grain and is 
eventually removed from the grain by radiative and 
evaporative heat loss is: 
\begin{equation}
\rho C_V {\partial T\over \partial t} = 
{\partial \over \partial x}\left(\kappa {\partial T\over \partial x}\right)
-g_r - g_e 
\end{equation}
where $x$ is the dimension along the linear chain. 
The terms on the RHS of the equation describe 
the roles of conduction and 
radiative and evaporative cooling, respectively. 
The heat conduction coefficient, $\kappa,$ 
is approximately temperature independent over the temperature 
range of interest and 
has been approximated by Tielens \& Allamandola (1987) as 
$\kappa = 0.3$ W K$^{-1}$ m$^{-1}$ (see also Appendix A).
The quantities 
$g_r$ and $g_e$ are the radiative and evaporative cooling rates 
per unit volume.

If we integrate over the volume of each subunit, the equation is then
\begin{equation}
V \rho C_V {\partial T\over \partial t} = 
\kappa A_{\rm eff} \ell\ {\partial^2 T\over \partial x^2} 
-\Gamma_r - \Gamma_e 
\end{equation}
where $V=(4\pi/3) r_s^3$ is the approximate volume of the subunit 
(the volume of the connection region between subunits is assumed to 
be negligible; see Appendix B), 
which may differ from the effective volume for the conduction term, 
$A_{\rm eff}\ \ell$.  In this expression, $A_{\rm eff} = \alpha r_s r_c$ 
is the effective area for conduction between subunits  
where $\alpha \approx 3$ (see Appendix B), and  
$\ell \simeq 2 r_s$ is the distance between thermal conduction surfaces. 
The quantities $\Gamma_r$ and $\Gamma_e$ are the radiative and 
evaporative cooling rates per subunit.  
The evaporative cooling rate is 
$\Gamma_e = (4\pi r_s^2/7$\AA$^2)\ \nu_0 E_b \exp(-E_b/kT)$ 
(e.g., Tielens \& Allamandola 1987) 
where $\nu_0= 2\times 10^{12}$ s$^{-1}$, 
the binding energy $E_b=960$ K for CO (Sandford \& Allamandola 1990), 
and the term $4\pi r_s^2/7$\AA$^2$ expresses the assumption 
that all available adsorption sites on the grain are filled. 
The radiative cooling rate is 
$\Gamma_r = 4\pi r_s^2\ \sigma T^4 Q_{\rm eff}$, where 
$Q_{\rm eff}\simeq 0.01 (r_s/1000$\AA$) 
(\lambda/100\mu {\rm m})^{-1}$
(e.g., Mezger et al.\ 1982). 

To determine the resulting evaporation rate, 
we discretized the thermal diffusion equation, representing 
each thermal subunit with a grid point. 
Since the equation is non-linear (e.g., the specific heat 
depends on temperature), 
we solved the discretized equation using an implicit 
method in which the temperature dependence of the heat capacity 
and the cooling terms were lagged.  
This approach produced rapid, stable solutions. 
For the end point conditions, we considered two possibilities.  
First, we considered the possibility that 
the ends of the chain are connected to a heat sink,  
i.e., the end points are highly conducting.  
This might be appropriate if the thermal 
subunits are thought to be part of a very large grain that has infinite 
capacity to absorb heat from the hot spot.  
Second, we considered the possibility that 
the ends of the chain are perfectly 
insulated so that there is no heat flux (${\partial T/\partial x} = 0$) 
through the end points. 
This would be appropriate if the chain of thermal subunits 
represents the entire grain. 

Figures 2--4 show the results for a representative case in which 
the central member of a linear chain of 151 subunits of 
radius $r_s=50$ \AA, with $r_c/r_s=0.02$, 
is hit by an X-ray that deposits 300 eV in the central subunit.  
All subunits are initially at 17 K, the assumed equilibrium grain 
temperature in the outer disk.  
(The results are insensitive to the value of the initial temperature. 
For example, essentially identical desorption rates are obtained with 
an initial temperature of 10 K.) 
The small value of $r_c$ is meant to represent the case in which the 
subunits are joined at their silicate cores rather than by their icy 
mantles. 
Figure 2a shows the time evolution of the temperature of the 
central subunit for a chain with conducting (solid line) and 
insulated (dotted line) ends.  
Figure 2b shows the total number of CO molecules evaporated as 
a function of time for the same cases. 
Figure 3 shows how the deposited energy is dispersed through 
the grain for both the conducting and insulated cases, 
and Figure 4 shows the time evolution of the temperature 
of selected individual thermal subunits in the insulated case.

As shown in Figure 2a, the hot spot reaches a fairly high temperature 
$> 140$ K. The end point condition makes little difference to the
temperature evolution of the hot spot until late in the evolution, at
$t=1 \mu$s.  At this time, the energy has diffused
throughout the entire chain (Fig.~4).  The insulated chain equilibrates to a
temperature $T > 17$ K, and the chain cools down radiatively beyond
that point.  In contrast, the chain with the conducting ends cools more
rapidly down to 17 K due to energy flow out of the chain (Figure 3).
In either case, most of the CO molecules are evaporated early on 
($t < 10$ ns) from the central subunit (Figure 2b).  Since the early time
evolution of the hot spot temperature does not depend on the end point
conditions, the evaporation rate is insensitive to the end point
conditions.  In either case, approximately 60\% of the deposited photon
energy goes into evaporation, and a negligible fraction of the
deposited energy is radiated during the interval in which evaporation
occurs.

We also explored the effect of a larger $r_c$, which might result 
if the subunits acquire an icy mantle before being joined. 
Figure 5 compares the results for $r_c/r_s=0.02$ and 0.1, 
with all other parameters the same as in Figures 2--4.
With a larger contact area, heat diffuses out of the grain 
more rapidly, reducing the evaporation rate. 
For $r_c=1$\AA\ and 5\AA, the fraction of the deposited energy 
that goes into evaporation declines from approximately 60\% to 
approximately 40\%. 

Alternatively, the effective conductivity between subunits might be
reduced due to several solid state effects outlined in greater detail
in Appendix A.  These include:  a decrease in the thermal conductivity
due to the small sizes of interstellar grains; the difference in the
composition of materials across a contact surface (e.g., at an
ice-silica transition); and the difficulty of propagating long
wavelength phonons through a relatively small interface.  Each of these
considerations leads to an effective reduction in the thermal
conductivity through the interface.
For example, at a given temperature, the dominant phonon frequency 
is $\nu_{\rm dom} = 4.25\ kT/h$.  For a Debye sound speed of 
$c_s=4.1\times 10^5 {\rm cm\,s^{-1}}$ (Pohl 1998),   
the corresponding dominant phonon wavelength is 
$\lambda_c = c_s/\nu_c$, or $\simeq 27$\AA\ at 17 K. 
This is much larger than the size of the contact area for 
$\sim$ 50\AA\ thermal subunits. 
To explore the effect of a reduced conductivity, 
Figure 5 also shows the thermal evolution for a chain with $r_s= 50$\AA\  
and $r_c/r_s=0.02$, in which the thermal conductivity is reduced to a 
fraction of its original value, i.e., $\kappa=f_k\kappa_0$, 
where $f_k=0.1$.  
The cool down of the hot spot is extended in this case, 
and the evaporation rate is consequently larger.  
Approximately 80\% of the deposited photon energy goes into 
evaporation.

Table 1 summarizes our results for 151-element chains with 
$r_s= 50$ and 100 \AA. 
For each $r_s$, the values of $E_d$ explored cover the likely range 
of deposited photon energies (Fig. 1 and \S 2.3).
As described above, for $r_s=50$\AA, X-ray desorption is efficient 
for $r_c/r_s=0.02-0.1$, and even more efficient when the effective 
$\kappa$ is reduced. 
X-ray desorption can also be efficient for $r_s=100$\AA, especially 
when the combination $f_k\ A_{\rm eff}$ is small. 
Note that the values of 
$N_{\rm des}$ tabulated here are less than that found for whole 
grain heating of grains $r_s$ in size (\S 2.3).  
With spot heating, some fraction of the energy leaks out of the 
hot spot and goes into raising the temperature of other subunits 
rather than toward desorption.  
In the whole grain heating case, we assumed that all 
the deposited energy in excess of that needed to raise the grain 
to 25 K was available for desorption.  These rates represent the 
maximal desorption rates, which would result if $f_k\ A_{\rm eff}$  
were appropriately small.

Thus, it appears that for spot heating to be efficient, the 
outer surfaces of large grains 
(i.e., where icy mantles are being accreted) must be
made up of regions $\sim 50-100$\AA\ in size that are poorly connected
thermally to the rest of the grain.  As discussed above, this is possible 
if large grains grow through the coagulation of much smaller grains.  
Such small grains are naturally produced by shattering processes in 
shocks.  
As described by Jones et al.\ (1996), shattering in shocks can convert 
$>40$\% of the grain mass into particles smaller than 100\AA.
The rapidity with which grains are shattered also implies that 
large grains must reform in the ISM, presumably through coagulation, 
creating porous, fractal grains. 
Given the steepness of the grain size distribution produced by shocks 
(steeper than MRN), it is even more likely that large grains, which 
may themselves be composite in nature, eventually 
come to be effectively ``coated'' with smaller grains.

In the picture of large grains coated by small grains, 
all grains are equal (independent of size) in terms of their rates of 
adsorption and desorption via spot heating.  That is, both the adsorption 
and spot heating desorption rates are proportional to grain surface area, 
and when the rates are equal for one grain size, they are equal for all 
grain sizes.  Consequently, the region of the disk surface over which 
a gas phase can be maintained by X-ray desorption is just a function of 
the X-ray irradiation rate seen by the grain (which causes desorption) 
and the local gas density (which regulates adsorption).

\section{Discussion} 

Two factors suggest the possible role of X-ray spot heating as 
an important physical process in outer protoplanetary disks: 
indications from disk chemistry studies that significant non-thermal 
desorption mechanisms are at work in disks at radii beyond 100 AU, 
and the possibly limited role of optical photon heating at 
these disk radii. 
The disk chemistry models seek to reproduce the molecular 
abundances measured at millimeter wavelengths in outer disks. 
In their study of the circumstellar dust and gas surrounding 
T Tauri stars, Dutrey et al.\ (1996) measured the $^{13}$CO and 
millimeter continuum emission in 33 systems.  Continuum emission 
was detected in $\sim 17$ systems, but $^{13}$CO was detected in only 
three systems.  In two of the three systems, the CO emission appeared to 
be more closely associated with outflows than with disks. 
Based on these data, Dutrey et al.\ 
concluded that CO is depleted by a factor $\sim 20$--100, most likely 
due to adsorption on grains.  The disk gas mass, 
as estimated from the CO emission, is typically 
less than $0.2 M_{\rm J}$ for T Tauri disks ($M_{\rm J}$ is the 
mass of Jupiter).  
In contrast, DM Tau was found to have a much larger 
inferred gas mass $\sim 1\ M_{\rm J}$, 
a result which may reflect its relative youthfulness.  

Aikawa \& Herbst (1999) have compared their predictions for 
disk chemical evolution with molecular line observations of 
DM Tau over disk radii 100 AU to 700 AU.  Even for this relatively 
gas rich system, they find good agreement with the observed 
molecular abundances if the sticking probability of all molecules 
to grains is 10 times smaller than typical estimates.  With 
conventional sticking probabilities (0.3--1.0; e.g., Williams 1993), 
CO is seriously depleted from the outer disk  
at ages of $3\times 10^5$ yr 
and densities $\gtrsim 10^5$ cm$^{-3}$. 
They interpret their result as an indication of the existence of a 
non-thermal desorption mechanism in disks.  

To estimate the efficiency of spot heating in maintaining a 
gas phase abundance of CO at large radii, we calculated the 
desorption rate given the assumptions made in \S 2.1 regarding 
the star and disk, and assuming that all grains are coated with grains of 
size $r_s=100$\AA.  In this case, the absorption of each photon with 
deposited energy $E_d>500$ eV may desorb somewhere between 
$\bar N_{\rm des} \sim 200-2200$ 
CO molecules given the range of values in Table 1. 
(The upper end of this range represents the possibility 
that we have overestimated the conduction rate between thermal 
subunits; see Appendix A).   
Therefore, the desorption rate is approximately 
\begin{equation}
D_X (a) \simeq  \bar N_{\rm des}(a)\ \int_{E_{\rm min}}^{E_{\rm max}} F_X(E)\ \sigma_X(a,E) \ dE.
\end{equation}
Comparing the resulting desorption rates with the adsorption rates 
calculated as in \S 2.3, we find that, at a disk radius of 300AU, 
desorption through X-ray spot heating is competitive with adsorption 
over the top $\Sigma_X = 0.002-0.006\gpersqc$ of the disk 
(or $\sim 2-4$\% of the total column density of the minimum mass solar 
nebula at that radius assuming irradiation of both disk surfaces).  
The range of column densities reflects the range 
$\bar N_{\rm des}=200-2200$. 
At 300 AU and a vertical column density of $\Sigma_X$, the 
gas density is $\sim 2\times 10^6$ cm$^{-3}$ and the 
column density along the line of sight to the X-ray source is 
$\sim 4-10$ times the vertical column density depending on the 
depth of the layer. 

The results are similar for grains coated with thermal subunits 
$r_s= 50$ \AA\ in size.  
In this case, the absorption of each photon with 
deposited energy $E_d>300$ eV may desorb somewhere between 
$\bar N_{\rm des} \sim 1500-3800$ 
CO molecules given the range of values in Table 1. 
As a result, at 300AU, 
desorption through X-ray spot heating is competitive with adsorption 
over the top $\Sigma_X = 0.004-0.006\gpersqc$ of the disk. 
Here, the higher desorption rate for the smaller thermal subunit 
compensates for the lower probability of X-ray energy absorption,  
resulting in a $\Sigma_X$ that is comparable to that found in the 
$r_s=100$ \AA\ case. 

The quantity $\Sigma_X$ is approximately constant with radius  
in the range 100 AU -- 300 AU because 
both the X-ray irradiation rate (and hence the desorption rate) 
and gas density in the surface layers (and hence the adsorption rate) 
decrease similarly with $r$. 
Thus, out to $\sim 300$ AU, the mass of CO that can be kept in the 
gas phase corresponds to a total mass of 
$2\Sigma_X \pi r^2 \simeq 0.1-0.4 M_{\rm J},$  
surprisingly close to the upper limits measured for gas masses in 
T Tauri disks (Dutrey et al.\ 1996). 
Similarly, the fractional column density of gaseous CO that can be 
maintained by X-ray spot heating (a few percent) is roughly 
consistent with 
the CO gas fractions that are observed in T Tauri systems.   

This agreement may be entirely fortuitous 
if one or more of our assumptions is in error.  
Perhaps most importantly, we have made a number of assumptions 
about grain substructure which could result in either an 
under- or overestimate of the efficiency of X-ray spot heating. 
On the one hand, we have assumed that large grains are completely 
coated by small grains which are poorly connected thermally to the 
rest of the grain.  The effective conductivity would be increased,  
and the desorption efficiency reduced, 
if the grain has undergone significant compaction. 
In addition, the efficiency of desorption by X-ray spot heating 
would be reduced if there is an incomplete coating of large grains 
by $\sim 50-100$\AA\ grains. 
Moreover, we have assumed that the icy mantles of grains are 
composed only of CO.  More likely, some of the X-ray energy deposited 
as heat would be expended desorbing $N_2$ and other molecules. 
On the other hand, we have been relatively conservative in our 
assumptions about several solid state effects that are 
relatively poorly understood from an astronomical perspective  
(\S 3 and Appendix A).
We have also assumed that the thermal subunits of grains are no smaller
than 50\AA.
In contrast, a variety of evidence suggests that the interstellar grain
size distribution extends significantly below 50\AA, down to the size of
molecular clusters (e.g., Desert, Boulanger \& Puget 1990;
Weingartner \& Draine 2001).
These smaller particles are likely to be heated to even higher
temperatures than found here through the absorption of X-ray and
possibly UV or optical photons.  Thus, by ignoring the possible 
excitation of smaller thermal subunits, we may underestimate the 
desorption rate.

We have also made some assumptions about the properties of disks 
and the limited role of other physical processes. 
We have assumed that a fairly substantial disk is present
at large radii, based on the extrapolation of the minimum mass solar
nebula, originally defined with in 40 AU, to large radii.  The predicted
limiting mass would be less if outer disks have much lower column 
densities than the extrapolated values.
We have also neglected to include the effect of photodissociation 
of CO at disk surfaces.  If we assume that CO is dissociated above a 
vertical column density of $7\times 10^{20}{\rm cm}^{-2},$ 
(van Dishoeck \& Black 1988), 
$\Sigma_X$ would be reduced by a factor of $\sim 2$.   
Note, however, that the H$_2$ shielding column above assumes the 
standard interstellar radiation field. 
The ambient UV field in the vicinity of protoplanetary disks may be 
reduced relative to this value by absorption in the foreground 
molecular cloud.  UV irradiation by the central star is another 
possibility, although its effectiveness may be limited since it must 
penetrate the line of sight column density to the star, which may be 
significantly larger than the corresponding vertical column density.  

Finally, desorption by X-ray spot heating should be considered 
in the context of optical photon heating, where issues such as 
grain substructure are not as critical.
The impact of stellar irradiation on the vertical structure of 
disks has been considered in a number of papers 
(e.g., Calvet et al.\ 1991; D'Alessio et al.\ 1998, 1999, 2001; 
Chiang \& Goldreich 1997, 1999). 
D'Alessio et al.\ (1998) assume that the gas and dust are well-mixed
(and at the same temperature)
and include the opacities from both sources in solving for the vertical
radiative transfer.  In contrast, Chiang \& Goldreich (1997)
show that gas-grain cooling is weak in the disk surface, and treat
the heating and reradiation of the dust independent of the gas.
These studies find that the absorption of stellar (UV to near-infrared)
photons in the disk atmosphere creates a super-heated surface
layer, a result that is due, in part, to the mismatch in the optical 
absorption and far-infrared emission efficiencies of grains.
The super-heated surface layer may be able to maintain a significant 
abundance of CO in the gas phase. 

In the formalism of Chiang and Goldreich (1997), stellar photons are 
absorbed in a surface layer that has a vertical optical depth in the 
visible of only $\tau = 0.4 R_*/r_d,$ or $5\times 10^{-5}$.  The corresponding 
vertical mass column density is $2\times 10^{-7} {\rm g\,cm^{-2}}$ 
if the dust in the upper disk atmosphere is similar to dust in the ISM. 
The energy deposited in the surface layer will warm the layers below. 
The extent of the warm region has been calculated by D'Alessio et al.\ 
(1998, 1999, 2001) in their studies of the vertical structure of 
disks around typical T Tauri stars ($L_\ast=1 L_\odot$, 
$M_\ast=0.5 M_\odot$). 
They find that irradiation produces a strong 
temperature inversion, with the column density and temperature of the 
heated layer depending on assumptions about the grain size 
distribution. 
Disks with an interstellar grain size distribution (50\AA\ to 1$\mu$m) 
are found to have a large temperature inversion layer and are too 
geometrically thick at large radii compared to observed disks around 
T Tauri stars (D'Alessio et al.\ 1999).
Disk atmospheres in which grain growth has occured (to 1 mm to 1 cm) 
are cooler, 
and therefore less geometrically thick.  They better reproduce the 
scale height and spectral energy distribution of T Tauri disks 
(D'Alessio et al.\ 2001).
Yet cooler disk atmospheres are found in the model of 
D'Alessio et al.\ (1998). 

What do these temperature distributions imply for the gas phase abundance 
of CO that can be maintained by thermal desorption?  
Since gas at 20 K and $n_{\rm H} \leq 10^7 {\rm cm}^{-3}$ 
(a typical density for the disk atmosphere at 200 AU) 
will experience little depletion of CO over the lifetime of 
T Tauri disks (e.g., Aikawa et al.\ 1996), 
we can estimate the column density of gas phase CO as the
disk column density that resides above 20 K. 
In the D'Alessio et al.\ (2001) models, 
e.g., for a power-law distribution of grain sizes $n\propto a^{-3.5}$ 
and a maximum grain size $a_{\rm max}=1$ mm, 
the disk surface temperature is 25 K at 200 AU and 
the disk column density 
that is above 20 K is $\sim 0.01\ {\rm g\, cm^{-2}},$  
i.e., a few times larger than the corresponding column density for 
X-ray spot heating.
In the D'Alessio et al.\ (1998) models, in which the outer disk 
atmosphere is cooler, a comparable column density is warmer than 
20 K much closer to the star, at $\simeq 90$ AU, where the 
disk surface temperature is $\sim 30$ K. 
Since the disk surface temperature decreases as $\propto r^{-2/5}$, 
even the disk surface is below 20 K at 250 AU.  
Extrapolating these results to disk radii $\sim$ 300 AU, it 
appears that, compared to X-ray spot heating, optical photon heating 
can maintain either a greater or smaller column density of 
gas phase CO depending on the details of the grain size distribution. 
The larger column densities of CO that appear to be predicted by 
the warmer disk models 
may in fact 
exceed the CO detection limits in typical T Tauri disks. 

If this is the case, there are several possible explanations for the 
discrepancy.  
One possibility is a higher grain albedo due to an icy grain mantle. 
In particular, if grain growth to millimeter sizes has 
occured in T Tauri disks (D'Alessio et al.\ 2001), the thickness of 
the ice mantle is likely to be significantly larger than the thickness 
of grain mantles in the ISM.  Thick ice mantles may result in a 
non-negligible grain albedo and a lower grain temperature.  The 
magnitude of the effect depends on the effective albedo of mantled 
grains, which appears to be uncertain.  The albedo of pure water ice 
is close to unity at optical wavelengths, but the albedo of dirty ice 
may be significantly lower.  Chiang \& Goldreich (1997) have previously 
estimated that an icy mantle may reduce the absorbed stellar flux by 
as much as 0.3 and, therefore, the grain temperature by 0.75. 
If the disk surface temperature in the D'Alessio et al.\ (2001) model 
is reduced by this factor, very little of the disk column density is 
above 20 K at radii $>$ 150 AU.

Given the above uncertainties in both grain thermal substructure and
the grain size distribution in T Tauri disks, it is difficult at
present to evaluate the effectiveness of X-ray spot heating from CO
observations of outer protoplanetary disks.  Although the chemical
robustness and low condensation temperature of CO makes it a popular
diagnostic of the physical properties of outer disks, CO may not be the
best diagnostic of the efficiency of processes such as X-ray spot
heating, in part, because other processes, such as thermal desorption,
may be competitive at low desorption temperatures.
Thus, based on models from the literature, we find that optical 
photon heating alone may be sufficient to explain the observed CO 
gas phase abundance of outer disks.  However, it also appears that 
X-ray spot heating alone may be efficient enough to explain the 
observations.

A more stringent test of the efficiency of X-ray spot heating in
protoplanetary disks could come from the study of molecules 
that are more strongly bound to the grains, and that consequently 
have higher evaporation temperatures. 
One example is CS, which has
an evaporation temperature of $\sim$40 K off of a pure 
silicate surface. 
To date CS emssion has been detected in two disks 
on size scales $\gtrsim 300$ AU 
(DM Tau---Dutrey et al.\ 1997; LkCa15---Qi 2001), 
suggesting that some non-thermal desorption mechanism is active 
at these radii.  
In a subsequent paper, we will examine
the efficiency of X-ray spot heating for CS and other molecular species.

\section{Summary} 

We have examined the ability of X-rays to desorb molecules 
from grains, in particular via X-ray spot heating, which may arise 
as a consequence of the aggregate nature of grains. 
A simple model of grain thermal substructure indicates that spot 
heating may be quite efficient at all grain sizes if large grains 
are coated with small ($\lesssim 100$ \AA) grains which are poorly 
connected thermally to the rest of the grain.  
Applying this model to outer protoplanetary disks, 
we estimate that for stellar
X-ray properties typical of T Tauri stars, X-ray spot heating can
sustain an equilibrium gas phase of CO over a vertical column density
of $\sim 0.002-0.006 \gpersqc$ at 300 AU, in fairly good agreement with
millimeter emission line observations of outer protoplanetary
disks.  Although spot heating, thus, appears to be a potentially
significant desorption mechanism, our current limited knowledge of
grain substructure and its consequences for thermal transport in
grains limits our ability to determine the true extent of its 
effectiveness.

Although we have discussed X-ray spot heating in the context of outer
protoplanetary disks, spot heating may also be a significant source 
of molecular desorption in other astrophysical contexts.  Since its 
effectiveness depends on the energy of individual photons rather than 
on photon flux, it may also provide a new mechanism for desorbing 
molecules at fairly large distances from X-ray emitting stars, 
e.g., in molecular clouds. 
Indeed, observations of both molecular clouds and outer protoplanetary 
disks, when compared with the predictions of future (more robust) X-ray 
spot heating models, may provide a means of constraining grain
substructure.

\acknowledgments 
We would like to thank Bruce Draine, Alexander Tielens, and 
Thomas Henning for insightful discussions and suggestions. 
This work was supported by a grant from 
NASA's Origins of Solar Systems program. 
The work of one of us (JNU) was performed under the auspices of 
the U.S. Department of Energy by the University of California, 
Lawrence Livermore National Laboratory under contract 
No. W-7405-ENG-48. 

\appendix
\section{Thermal Properties of Interstellar Aggregate Analogs}

Most of the silicates in the interstellar medium are believed to be
amorphous in form (e.g. Tielens \& Allamandola 1987), although there
may be a small crystalline component (Demyk et al. 1999; Cesarsky et
al. 2000).  For heat transport the thermal properties of interest are
the specific heat at constant volume, $C_v$ (or pressure: $C_p$) and
the thermal conductivity ($\kappa$).  To estimate these quantities the
most commonly used ``analog''  for interstellar silicates is amorphous
(vitreous) silica (e.g. Leger et al. 1985).

Since the examination of the specific heat and conductivity of vitreous
silica by Zeller \& Pohl (1971) there have been significant advances in
the characterization of the thermal properties of amorphous
substances.   Of particular note for interstellar studies is the near
universal thermal conductivity and specific heat of amorphous solids
(see review by Pohl 1998).  Simply put, all amorphous solids studied
(at present) show similar values of $\kappa$ and $C_p$, to within an
order of magnitude, for a wide range of temperatures, 0.1 K $<$ T $<$
1000 K despite the variation in both quantities with 
temperature.   The reason for the near universality is believed to be
related to the disordered state of an amorphous solid and is a focus of
current solid state research (Pohl 1998).  From an astronomical point
of view, this similarity is encouraging because we can plausibly model 
heat diffusion in amorphous interstellar grains without an exact knowledge 
of their composition.  For
this study we have used the fit provided to the specific heat of
amorphous silica by Leger et al. (1985).  This fit is valid for the
temperature regimes of interest in our study (T $<$ 200 K), but is not
valid for greater temperature.  For the thermal conductivity we have
used the (constant) value given by Tielens \& Allamandola (1987), 
which is a reasonable approximation of the thermal conductivity coefficient 
(see Zeller \& Pohl 1971) over the temperature range of interest. 

The small size of interstellar grains, compared to the typical sizes 
of laboratory samples, can have important
consequences for heat retention and diffusion.  
There are strong mathematical and intuitive analogies between 
the behavior of elastic waves in matter (phonons) and electromagnetic 
waves. In particular, the finite size of a solid body constrains 
the allowed phonon wavelengths in the same way that a waveguide 
constrains allowed electromagnetic modes.  In analogy to heat
transport in a classical gas, the thermal conductivity of a solid can be
expressed as:

\begin{equation}
\kappa = \frac{1}{3} C_v \bar{v} l
\end{equation}

\noindent where $\bar{v}$ is the mean velocity of phonons 
in the solid and $l$ is the phonon mean free path 
(e.g. Davies 1978).  The mean velocity ($\bar{v} = 4.1 \times 10^5$ cm
s$^{-1}$; Cahill, Watson, \& Pohl 1992) and specific heat are measured
quantities, and at T = 10 K the mean free path inferred from 
measurements is $\sim 1000$\AA .  
As a result, the adopted thermal conductivity of amorphous silica
(determined by Zeller \& Pohl in a sample size of 5 $\times$ 5 $\times$
40 mm) could be severely reduced in sample sizes $\lesssim$100 \AA .  
The decrease in thermal conductivity with crystal size is well known 
(e.g. Thacher 1967) and has even been examined in the context of the 
Martian surface (Presley \& Christensen 1997).  
Nevertheless, the exact dependence of the thermal conductivity  
on size remains uncertain, especially at size scales $\lesssim 100$\AA.  
Due to the enforced reduction in the mean free path,
eq.\ A1 suggests a linear dependence, but some studies 
suggest both a size (square-root of the particle size)
and pressure dependence (Presley \& Christensen 1997).
Given the possibility that the thermal conductivity decreases  
significantly at small size scales, we probably underestimate the 
heat retention and evaporative cooling rate of grains. 

This can have even more extreme effects if we consider a porous
aggregate composed of particles connected by a small and finite contact
area.  Since the contact radius through which conduction would occur is
smaller than the radius of either of the conjoined  particles (see
\S 3), it can be expected that phonon propagation
will be curtailed to some extent through the connection (as opposed to
within a single particle).   In our model of an aggregate composed of
50 \AA\ or 100 \AA\ particles\footnote{ We note that these radii are
reasonable when compared to the smallest grains used in size
distributions determined by comparison to the UV extinction curve (25
\AA: Kim, Martin \& Hendry 1994; 10 \AA: Zubko, Krelowski, \& Wegner
1996).  In addition, very small grains are inferred to exist in the 
ISM based on short wavelength infrared (3--60$\mu$m) emission from
interstellar dust (e.g., Weingartner \& Draine 2001).  Thus, our 
assumption that grains are aggregates of particles 50--100\AA\ in size 
is conservative if aggregate grains include a substantial population 
of grains smaller than 50\AA.}, the contact radius is between 
2--10\% of the particle size.
This contact radius is much less than the dominant phonon wavelength.  
The dominant phonon frequency can be expressed in a fashion similar 
to Wien's Law (see Cahill \& Pohl 1988):

\begin{equation}
\nu_{dom} = 4.25 \frac{k}{h}T.
\end{equation}

\noindent At 10 K the dominant wavelength ($\lambda_{dom} =
\bar{v}/\nu_{dom}$) is $\sim$50 \AA.
We can expect that the phonons with wavelengths 
much greater than the contact radius will not participate in heat
diffusion throughout the aggregate as a whole, with the net result
being a reduction in the thermal conductivity through the contact
point.  

A final related effect that must be noted in regard to heat diffusion
in interstellar aggregates is thermal boundary resistance (also known
as Kapitza resistance).  This ``resistance'' to heat diffusion occurs
at the interface between two dissimilar solids and can lead to a
temperature discontinuity at the boundary (Little 1959; Swartz \& Pohl
1989).  The relevant boundary for interstellar grains would be between
silicates and water ice -- differences in material density and sound 
velocity across this boundary will lead to some phonons being reflected 
and others transmitted, with the net result again being an increase 
in the rate of evaporative cooling.

\section{Effective Area for Conduction Between Thermal Subunits}

The effective area for conduction between thermal subunits, $A_{\rm eff},$ 
is estimated assuming that the chain of subunits is made of spherical 
subunits of radius $r_s$ that are connected to one another by 
cylindrical connecting volumes (by ``necks'') of radius $r_c$ and 
half-length $z_c$.  To match the sphere to the cylinder, the sphere is 
truncated near the connecting points, at a distance $r_t$ from 
the center of the sphere so that the area of the truncation region 
equals the crosssectional area of the cylinder 
(see, e.g., Chokshi et al.\ 1993, Fig.~2).  

For this geometry, the geometric part of the thermal impedance of 
the truncated sphere is obtained by integrating
$$I_s = \int_0^{r_t}  {dx\over \pi(r_s^2 - x^2)} = 
	{1\over 2\pi r} \ln \left\vert{(r_s+r_t)\over (r_s-r_t)}\right\vert $$ 
where $\pi(r_s^2 - x^2)$ is the crosssectional area of the sphere 
at a distance $x$ from the center of the sphere.
The geometric part of the thermal impedance of the half-cylinder is 
$I_c = z_c/\pi r_c^2$. 
Since $z_c$ is likely to be 1--10\AA\ (e.g., Chokshi et al.\ 1993), 
comparable to our assumed values of $r_c$ for $r_s=50-100$ \AA,  
we assume $z_c=r_c$. 
The effective area is, therefore, $A_{\rm eff} = (r_t + z_c)/(I_s+I_c).$  
As an example, for a sphere of radius $r_s=1$, that is matched  
to a cylinder of radius $r_c=0.02$, 
the effective area of truncated sphere plus cylinder is 
$A_{\rm eff} = 0.0587$, or $A_{\rm eff}=\alpha r_c r_s$ where 
$\alpha \approx 3$.

\vfill\eject

\begin{figure}
\plotone{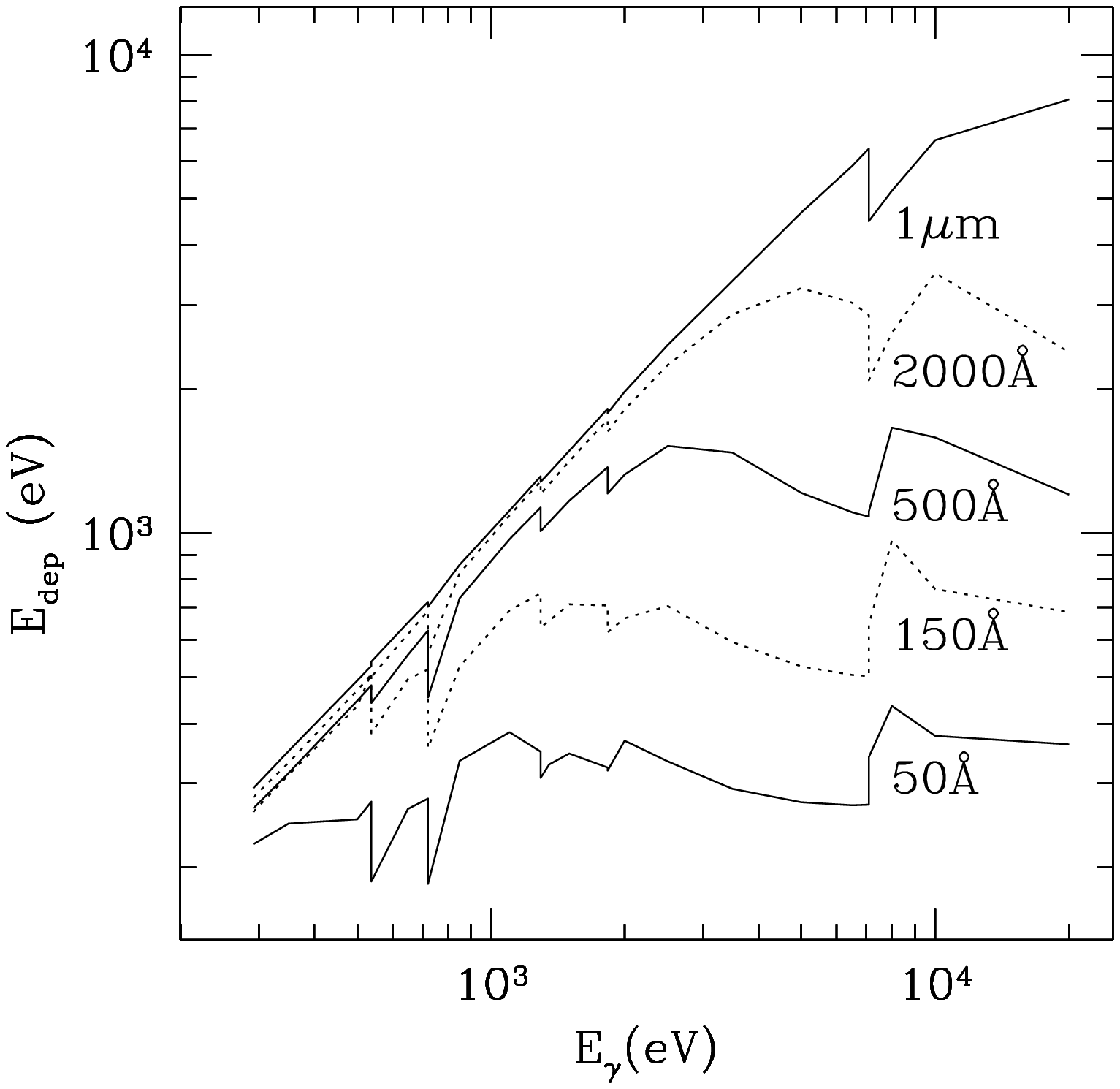}
\caption{Deposited energy as a function of incident X-ray energy.
\label{Edep}
}
\end{figure}

\begin{figure}
\plotone{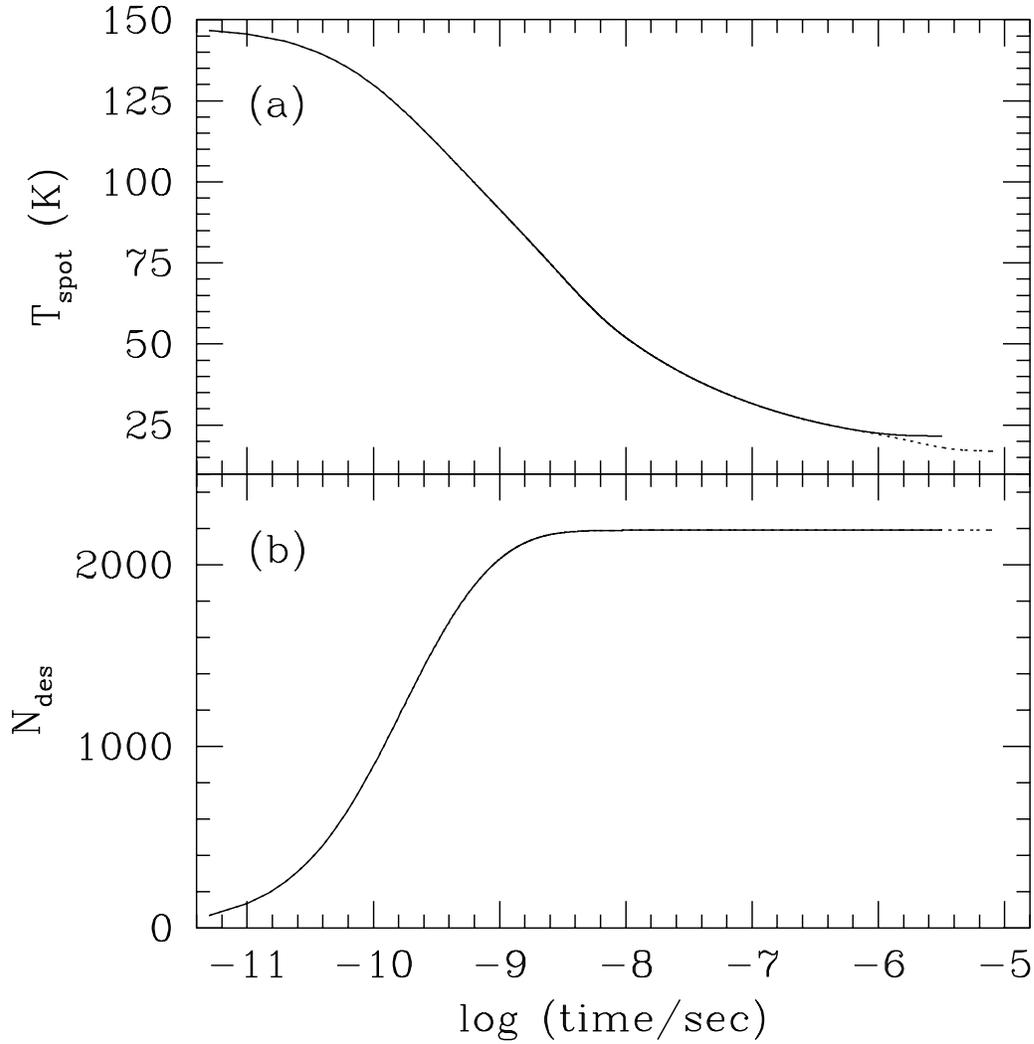}
\caption{Results for the case $r_s =50$ \AA, $r_c/r_s=0.02$, 
and $E_d = 300$ eV, for a 151 subunit chain with insulated (solid line) 
and conducting (dotted line) ends:
(a) time evolution of the temperature of the hot spot, and 
(b) total number of CO molecules evaporated as a 
function of time.  
\label{Tnhot}
}
\end{figure}

\begin{figure}
\plotone{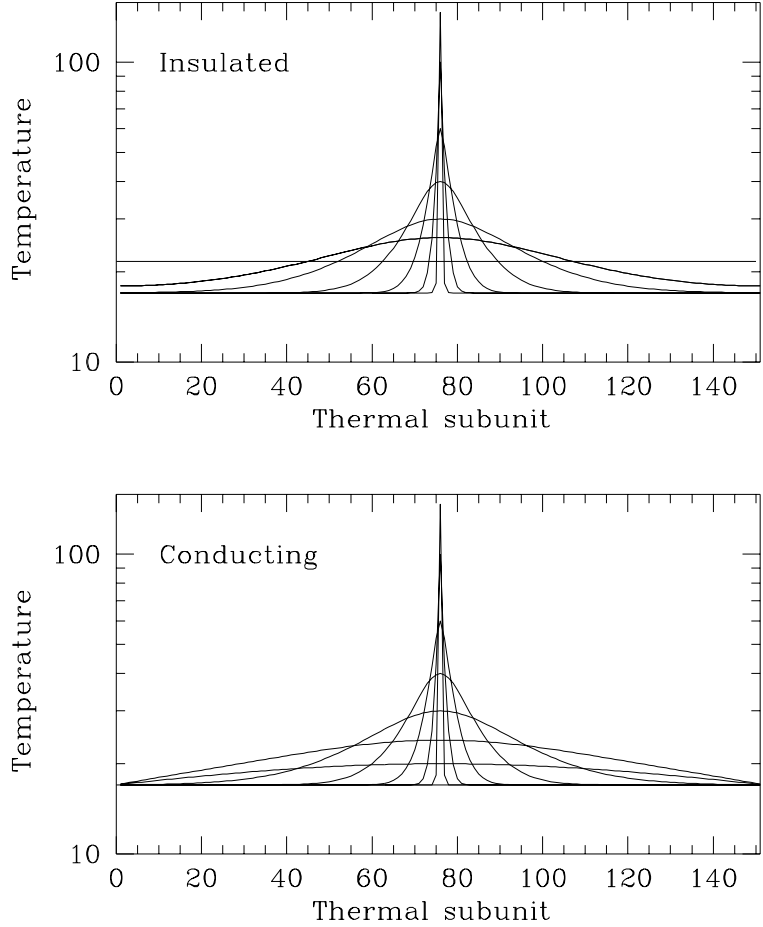}
\caption{Time evolution of the grain thermal substructure 
for the case $r_s =50$, $r_c/r_s=0.02$, and $E_d = 300$ eV, for a 
151 subunit chain with insulated ({\it top}) or conducting ({\it bottom}) 
ends.  
In the insulated case, the temperature structure is shown at time 
intervals 
$5\times 10^{-12}$ s, 
$6\times 10^{-10}$ s, 
$6\times 10^{-9}$ s, 
$3\times 10^{-8}$ s, 
$1\times 10^{-7}$ s, 
$3\times 10^{-7}$ s, and  
$2\times 10^{-6}$ s. 
In the conducting case, the temperature structure is shown at time 
intervals 
$5\times 10^{-12}$ s, 
$6\times 10^{-10}$ s, 
$6\times 10^{-9}$ s, 
$3\times 10^{-8}$ s, 
$1\times 10^{-7}$ s, 
$6\times 10^{-7}$ s, 
$2\times 10^{-6}$ s, and 
$8\times 10^{-6}$ s. 
\label{thermalsubstructure}
}
\end{figure}

\begin{figure}
\plotone{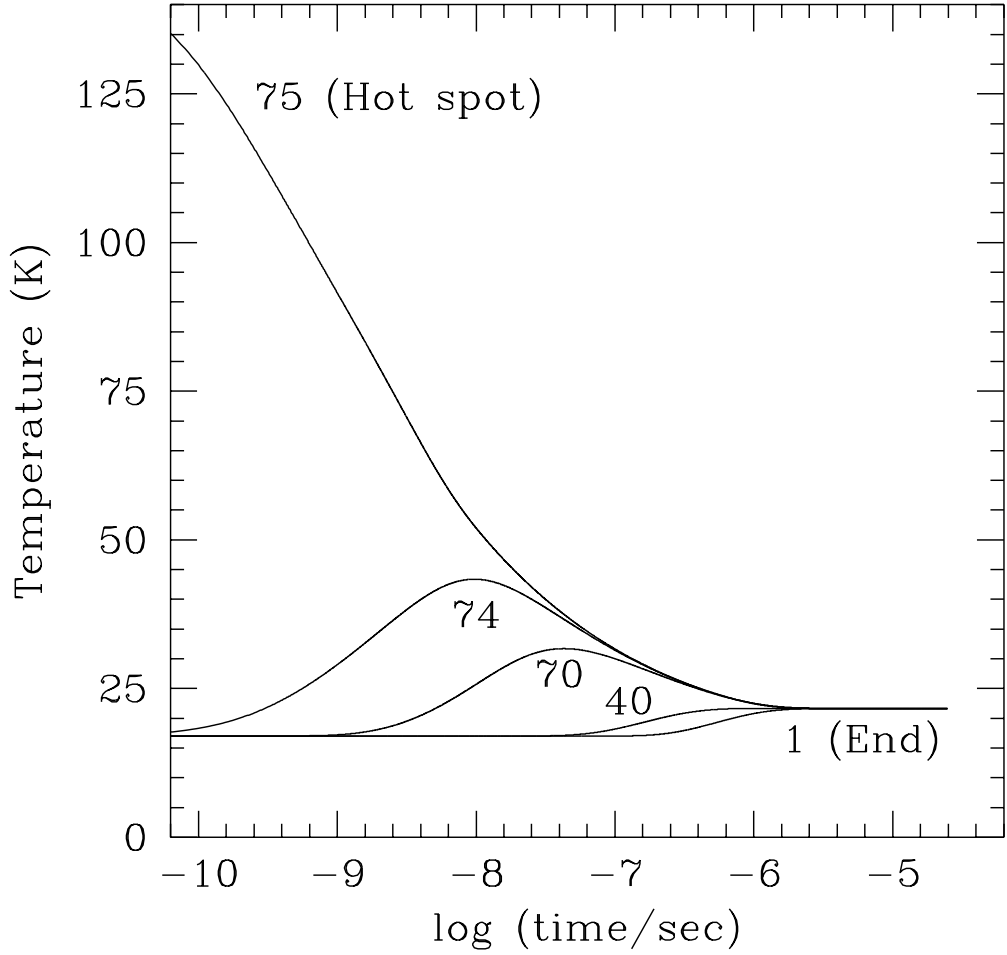}
\caption{Time evolution of the temperature of selected individual 
thermal subunits in the insulated case shown in Figure 3.  
The time evolution of the temperature at the end of the chain 
(the first element of the chain), at the hot spot (the 75th element 
of the chain), and at intermediate locations in the chain are shown. 
\label{thermalevolution}
}
\end{figure}

\begin{figure}
\plotone{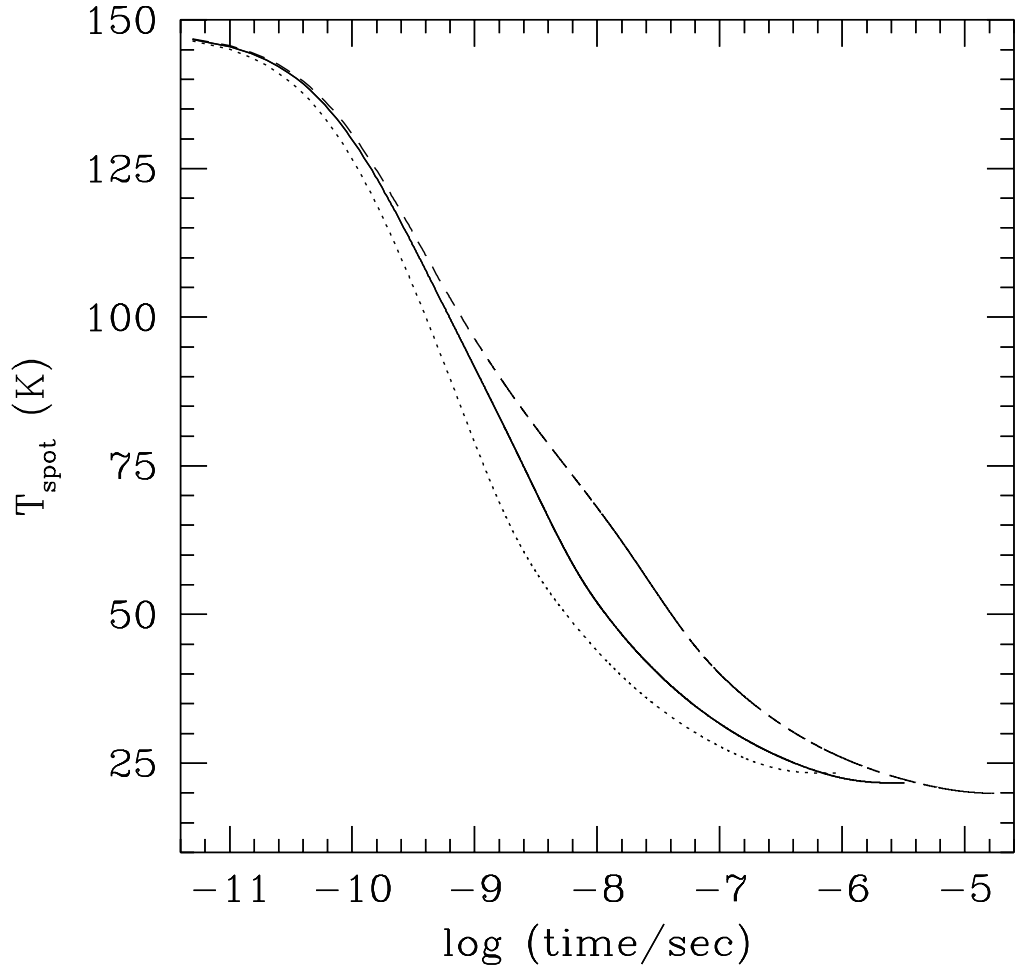}
\caption{Time evolution of the temperature of the central subunit 
for the case $f_k=1,$ $r_s=50$\AA, and 
$r_c/r_s=0.02$ (solid line) or $r_c/r_s=0.1$ (dotted line).  
For comparison, the case 
$r_s=50$\AA, $r_c/r_s=0.02$, and $f_k=0.1$ (dashed line) is also shown. 
\label{TnhotA}
}
\end{figure}
\clearpage

\begin{table}
\begin{center}
\centerline {Table 1. CO Desorption from Spot Heating}
\begin{tabular}{rrrrrr}
\tableline\tableline
$r_s$   &  $E_d$(eV) &  $r_c/r_s$  &  $f_k$  &  $T_{\rm spot}$  &  $N_{\rm des}$ \\ 
\tableline
50\AA\  & 300   & 0.02  & 1.0   & 147   & 2190 	\cr 
        &       &       & 0.1   & 147   & 2820 	\cr 
        &       & 0.1   & 1.0   & 147   & 1520 	\cr 
        &       &       & 0.1   & 147   & 2460 	\cr 
50\AA\  & 380   & 0.02  & 1.0   & 162   & 3130 	\cr 
        &       &       & 0.1   & 162   & 3790 	\cr 
        &       & 0.1   & 1.0   & 162   & 2380 	\cr 
        &       &       & 0.1   & 162   & 3420 	\cr 
\tableline
100\AA\ & 400   & 0.02  & 1.0   &  71   &   10 	\cr 
        &       &       & 0.1   &  71   &  120 	\cr 
        &       & 0.1   & 1.0   &  71   &    0 	\cr 
        &       &       & 0.1   &  71   &   30 	\cr 
100\AA\ & 500   & 0.02  & 1.0   &  78   &  240 	\cr 
        &       &       & 0.1   &  78   & 1250 	\cr 
        &       & 0.1   & 1.0   &  78   &   60 	\cr 
        &       &       & 0.1   &  78   &  470 	\cr 
100\AA\ & 600   & 0.02  & 1.0   &  83   &  580 	\cr 
        &       &       & 0.1   &  83   & 2210 	\cr 
        &       & 0.1   & 1.0   &  83   &  160 	\cr 
        &       &       & 0.1   &  83   & 1030	\cr 
\tableline
\end{tabular}
\end{center}
\end{table}

\end{document}